\documentclass[12pt]{article}
\usepackage{graphicx}
\usepackage[subrefformat=parens,labelformat=parens]{subfig}
\usepackage{hyperref}
\usepackage{amsmath}

\newcommand\pubnumber{DPF2015-360}
\newcommand\pubdate{\today}

\def\napoli{Department of Physics, McGill University, Montreal QC, Canada}

\def\Title#1{\begin{center} {\Large #1 } \end{center}}
\def\Author#1{\begin{center}{ \sc #1} \end{center}}
\def\Address#1{\begin{center}{ \it #1} \end{center}}

\newcommand\pubblock{\rightline{\begin{tabular}{l} \pubnumber\\
         \pubdate  \end{tabular}}}
\newenvironment{Abstract}{\begin{quotation}  }{\end{quotation}}
\newenvironment{Presented}{\begin{quotation} \begin{center} 
             PRESENTED AT\end{center}\bigskip 
      \begin{center}\begin{large}}{\end{large}\end{center} \end{quotation}}

\begin{document}
\begin{titlepage}
\pubblock

\vfill
\Title{The updated ATLAS Jet Trigger for the LHC Run~II}
\vfill
\Author{Sebastien Prince}
\Address{\napoli}
\Author{on behalf of the ATLAS collaboration}
\vfill
\begin{Abstract}
After the current shutdown, the LHC is about to resume operation for a new data-taking period, when it will operate with increased luminosity, event rate and center of mass energy. The new conditions will impose more demanding constraints on the ATLAS online trigger reconstruction and selection system. To cope with such increased constraints, the ATLAS High-Level Trigger, placed after a first hardware-based Level~1 trigger, has been redesigned by merging two previously separated software-based processing levels. In the new joint processing level, the algorithms run in the same computing nodes, thus sharing resources, minimizing the data transfer from the detector buffers and increasing the algorithm flexibility.

The jet trigger software selects events containing high transverse momentum hadronic jets. It needs optimal jet energy resolution to help rejecting an overwhelming background while retaining good efficiency for interesting jets. In particular, this requires the CPU-intensive reconstruction of tridimensional energy deposits in the ATLAS calorimeter to be used as the basic input to the jet finding algorithms. To allow this costly reconstruction step, a partial detector readout scheme was developed, that effectively suppresses the low activity regions of the calorimeter and significantly reduces the needed resources. In this paper we describe the overall jet trigger software and its physics performance. We then focus on detailed studies of the algorithm timing and the performance impact of the full and partial calorimeter readout schemes. We conclude with an outlook of the jet trigger plans for the next LHC data-taking period.
\end{Abstract}
\vfill
\begin{Presented}
DPF 2015\\
The Meeting of the American Physical Society\\
Division of Particles and Fields\\
Ann Arbor, Michigan, August 4--8, 2015\\
\end{Presented}
\vfill
\end{titlepage}

\section{Introduction}
The CERN Large Hadron Collider (LHC) \cite{LHC} was in shutdown in 2013 and 2014. Such a downtime period, separating the data-taking periods known as Run~I and Run~II, was required to perform the maintenance and upgrade work necessary to increase the collision energy, luminosity and bunch frequency that the LHC can achieve.

For the LHC detectors, such as ATLAS~\cite{ATLAS}, the increase in energy and luminosity means that there is be an increase by approximately a factor five of interesting collisions. One of the challenges for the ATLAS detector is to be able to record this higher production of interesting events while maintaining a good efficiency at low energy. An additional challenge comes from the expected higher number of interactions per bunch crossing, called pileup. This higher pileup increases the detector occupancy and thus the time required to reconstruct an event. The challenge is, in other words, to control the increased trigger rate with a limited latency.

The ATLAS trigger~\cite{Trigger} is the system that decides, in real-time, whether to record or not an event. Its main purpose is to discard the least interesting events to reduce its input frequency of 20~MHz (40~MHz), the bunch crossing frequency of the Run~I (Run~II) LHC, to an achievable recording rate of the order of 400~Hz (1~kHz). To accomplish this, the ATLAS trigger system is divided into consecutive levels: Level~1, implemented in hardware, and a High-Level Trigger (HLT), implemented in software~\cite{TDAQ1}. Each level allows the subsequent ones to have a longer latency, and thus more refined selection algorithms.

As the LHC is a hadronic collider, colored particles are the most prevalent high-energy particles produced. Due to color confinement, these particles shower and hadronize to form collimated sprays of particles, called jets. These particle jets are detected as energy deposits in the electromagnetic and hadronic calorimeters. However, the ATLAS jets are reconstructed from all calorimetric energy deposits, including those associated to non colored particles. The main increase in the jet rate comes from low-energy objects, since their high cross-section increases further with the higher collision energy. Also, the rate of jets not coming from the hard-scattering process, considered as background, increases as the pileup does. Therefore, the trigger system specialized in selecting jets, the jet trigger, had to undergo major changes to maintain a good selection performance in Run~II.

\section{Jet trigger performance in Run~I}
In Run~I, the HLT was divided into two software levels: the Level~2 and the Event Filter. The algorithms used by the jet trigger to reconstruct jets were different across the three trigger levels. They were chosen such that they maximize the performance within the available latency. At all levels, the primary selection criterion of the jet triggers is on the transverse energy of the jets built by the algorithms. 

At Level~1, the algorithm was a sliding window~\cite{L1} with trigger towers as input, built across the whole calorimeters. Trigger towers are the sum of cells along the depth of the calorimeters, at a specific angular\footnote{ATLAS uses as angular variables the azimuthal angle $\phi$ and the pseudorapidity $\eta=-\ln\tan\frac{\theta}{2}$, where $\theta$ is the polar angle.} position. The position of the sliding window at a local energy maximum is called a region of interest (RoI)~\cite{ROI}.

At Level~2, two algorithms were used, depending on whether cells or trigger towers were used as input. If the input were calorimeter cells,  the algorithm was a three-iteration cone algorithm~\cite{Cone} with a starting axis pointing at the RoI. Only cells around the RoI were taken as input. Instead, if the input were trigger towers, the algorithm was instead the same collinear- and infrared-safe algorithm that is used for offline jet reconstruction: anti-$k_t$~\cite{Antikt}. The input trigger towers were not restricted to be near the RoI. Since the trigger towers are less granular than cells, the timing of the algorithm still satisfied the Level~2 latency requirement even without any angular distance restriction.

At the Event Filter, anti-$k_t$ was also used but now taking as input topological clusters over the whole calorimeters, as in the offline jet reconstruction. Topological clusters are noise-suppressed objects formed by merging in three dimensions neighboring calorimeter cells that are above some noise thresholds~\cite{Topoclusters}.

\begin{figure}[htb]
\centering
\subfloat[]{\includegraphics[height=2in]{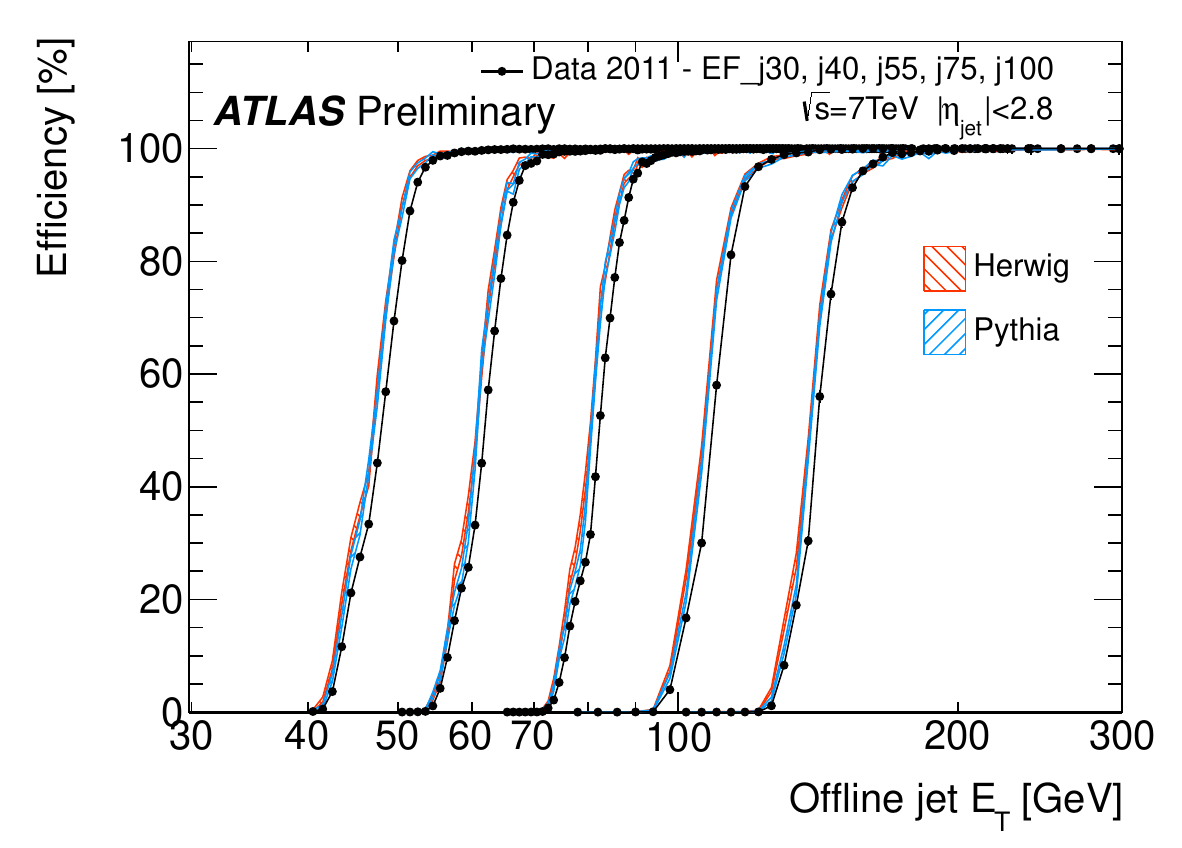}\label{fig:EF}}
\,
\subfloat[]{\includegraphics[height=2in]{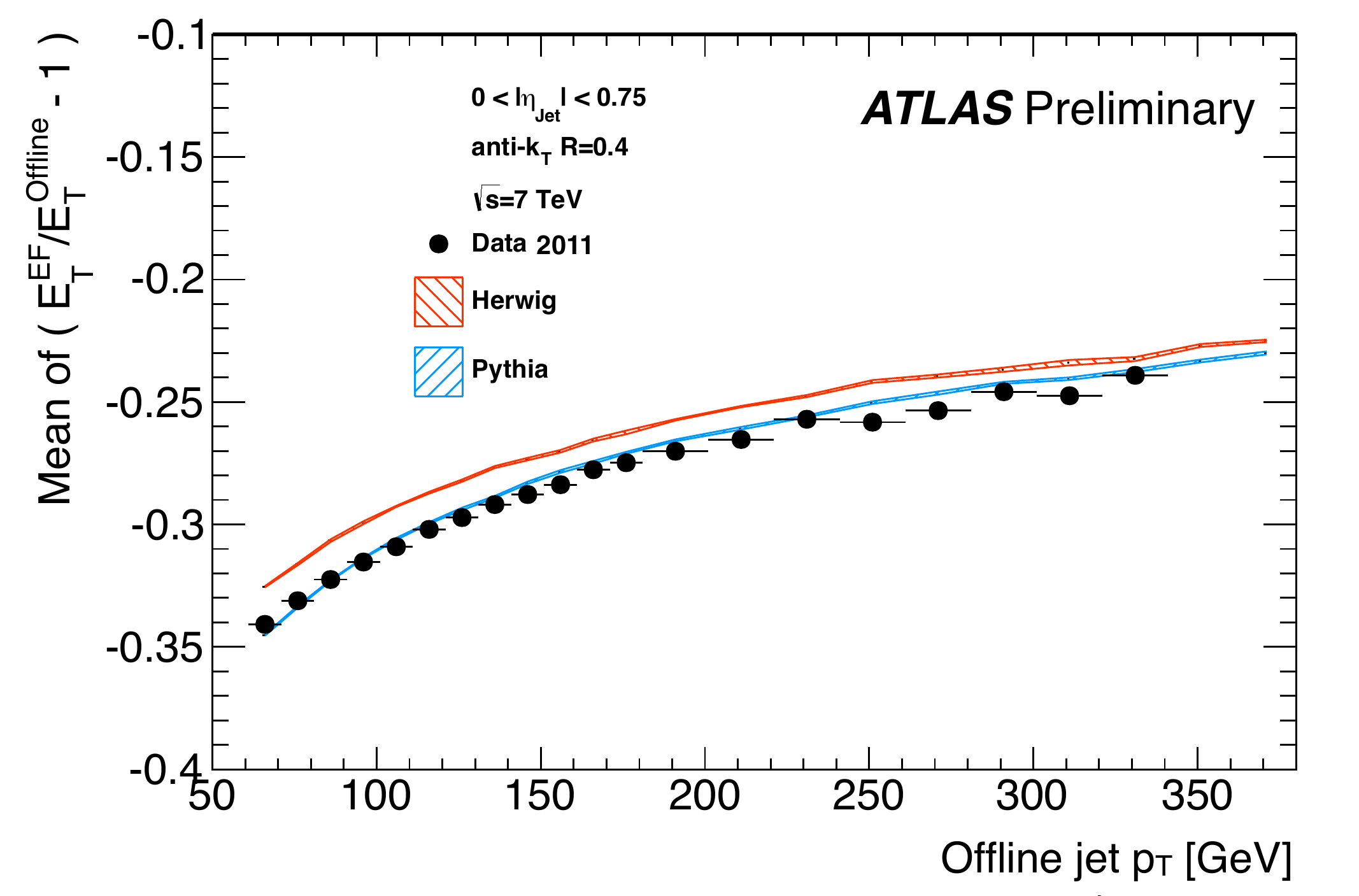}\label{fig:Offset}}
\caption{For data and Monte Carlo simulations, (a) the Event Filter trigger efficiency curves and (b) their offset relative to the offline $E_\text{T}$~\cite{JetTriggerPublic}.}
\label{fig:RunI}
\end{figure}

The performance of the Event Filter decision can be investigated from the trigger efficiency, shown in Fig.~\subref*{fig:EF}. The trigger efficiency is the ratio of the number of jets reconstructed online with respect to the number of offline-reconstructed jets. The plateau region reaches 100\% efficiency and the data agree with the Monte Carlo simulation in that region, showing that the trigger had a good performance. As can be seen, the turn-on region is not centered on the threshold value: in the case of EF\_j75 (an Event Filter trigger requiring a jet with at least 75 GeV of transverse energy), it is around 110~GeV. Figure~\subref*{fig:Offset} shows that relative difference between the energy calculated online and offline, across the full $p_\text{T}$ range. This discrepancy is due to the usage of different jet energy scales. Indeed, the offline reconstruction permits the application of a more precise calibration~\cite{JES}, while the short latency of the trigger system didn't allow for such detailed corrections.

\section{Jet trigger improvements for Run~II}
To be able to record the higher rate of interesting events in Run~II, the ATLAS trigger system received a multitude of upgrades~\cite{TDAQ2} during the long shutdown. An important change relevant to the jet trigger is the merging of the Level~2 and the Event Filter, such that the HLT became only one trigger level, effectively pooling the computing resources of the two levels and minimizing data transfers. The merged HLT allows the CPU-intensive step of building topological clusters for all events passing the Level~1. As the topological clusters are inherently noise-suppressed, they provide higher resolution than other types of input. Another important upgrade to the trigger system relevant to the jet trigger is the new faster readout boards that allow accessing the cell information across the whole calorimeters more frequently. In practice, this means being able to build topoclusters over the whole calorimeters for all events passing first trigger level, implying that the Level~2 algorithms can be skipped altogether. Therefore, in Run~II, the jet trigger algorithms are a sliding window over trigger towers at Level~1 and anti-$k_T$ over topological clusters at the HLT, both types of input being taken over the whole calorimeters.

Although simulations showed that this new design would respect the latency constraints of the online environment, a fallback plan was still conceived: a partial scan of the calorimeters~\cite{PartialScan}. The partial scan is an improvement over the algorithms limited to the information near the RoIs as it accesses all the RoIs simultaneously. This avoids processing multiple times energy deposits in regions covered by more than one RoI, while still suppressing the low activity regions of the calorimeters.

\begin{figure}[htb]
\centering
\subfloat[]{\includegraphics[height=1.7in]{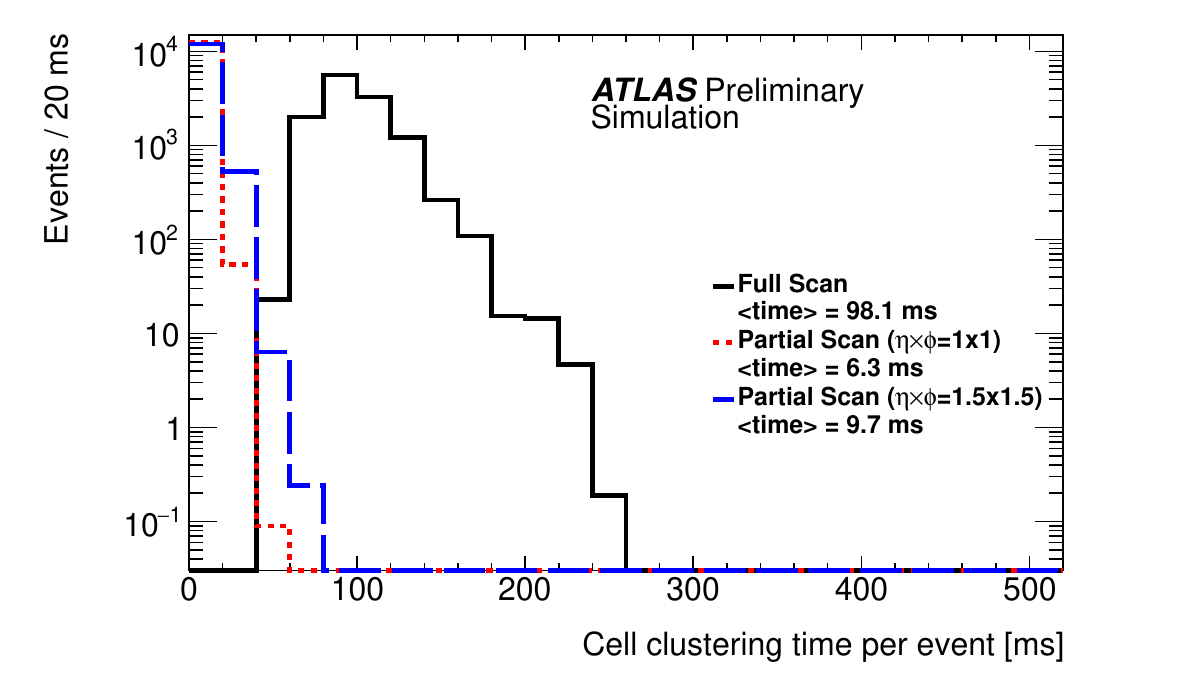}\label{fig:PSTime}}
\,
\subfloat[]{\includegraphics[height=1.7in]{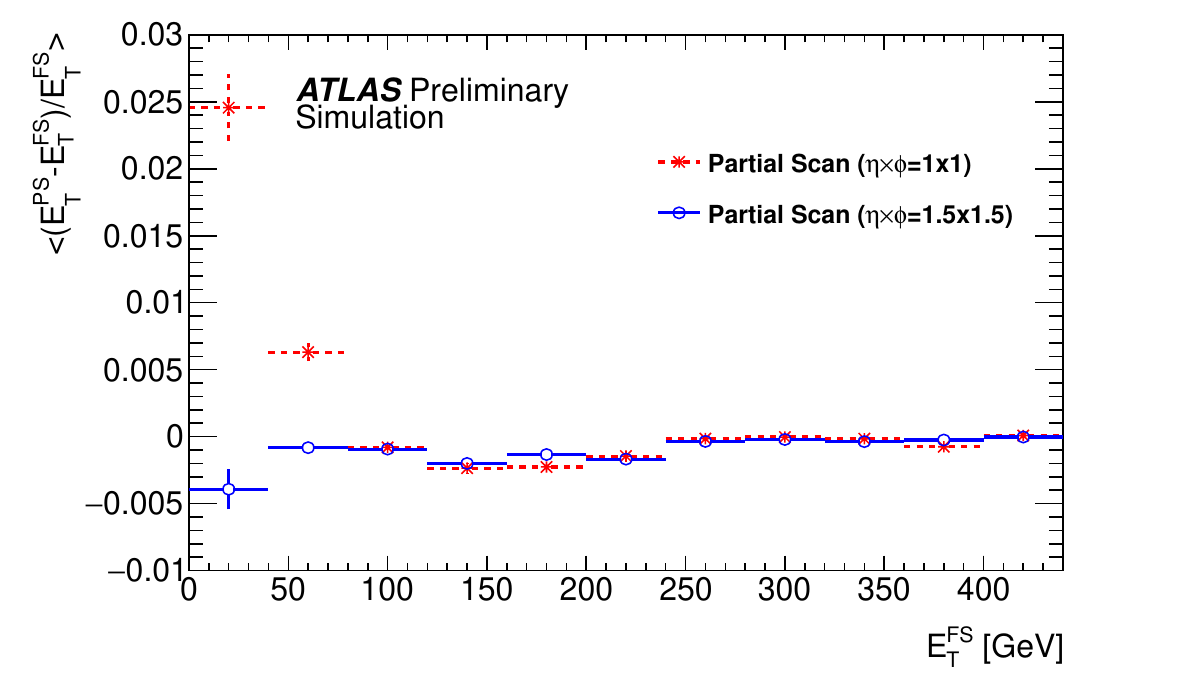}\label{fig:PSDiff}}
\caption{Comparison in simulation of the performance of partial scan, for two sizes of the region of interest, against full scan (a) for the topological clustering timing and (b) for the jet transverse energy measurement~\cite{PartialScan}.}
\label{fig:PS}
\end{figure}

Figure~\subref*{fig:PSTime} shows that, in simulation, the partial scan indeed reduces the timing of building topological clusters on average from 6\% to 10\% of that of the full calorimeter scan, depending on the size of the RoI. Where the partial scan falls short however is in the jet energy measurement. Figure~\subref*{fig:PSDiff} shows the relative measured energy difference between the partial scan and the full scan techniques. Although at higher $E_\text{T}$ the difference is small, it is non negligible at lower $E_\text{T}$, where the production rate of jets is higher. This is mitigated by using larger RoIs, but a difference still remains. For this reason, as long as the trigger decision can be achieved within the HLT latency when using a full scan of the calorimeters, such a technique is to be preferred.

\section{Jet trigger performance in Run~II}
To implement the new design of the jet trigger within the merged HLT, the jet trigger software had to be completely rewritten. This was taken as an opportunity to streamline the software, which would facilitate implementing further refinements. Indeed, a streamlined code allowed to implement some of the offline jet energy corrections in the trigger: the pileup subtraction and the jet energy scale calibration~\cite{JES}.

\begin{figure}[htb]
\centering
\includegraphics[height=2in]{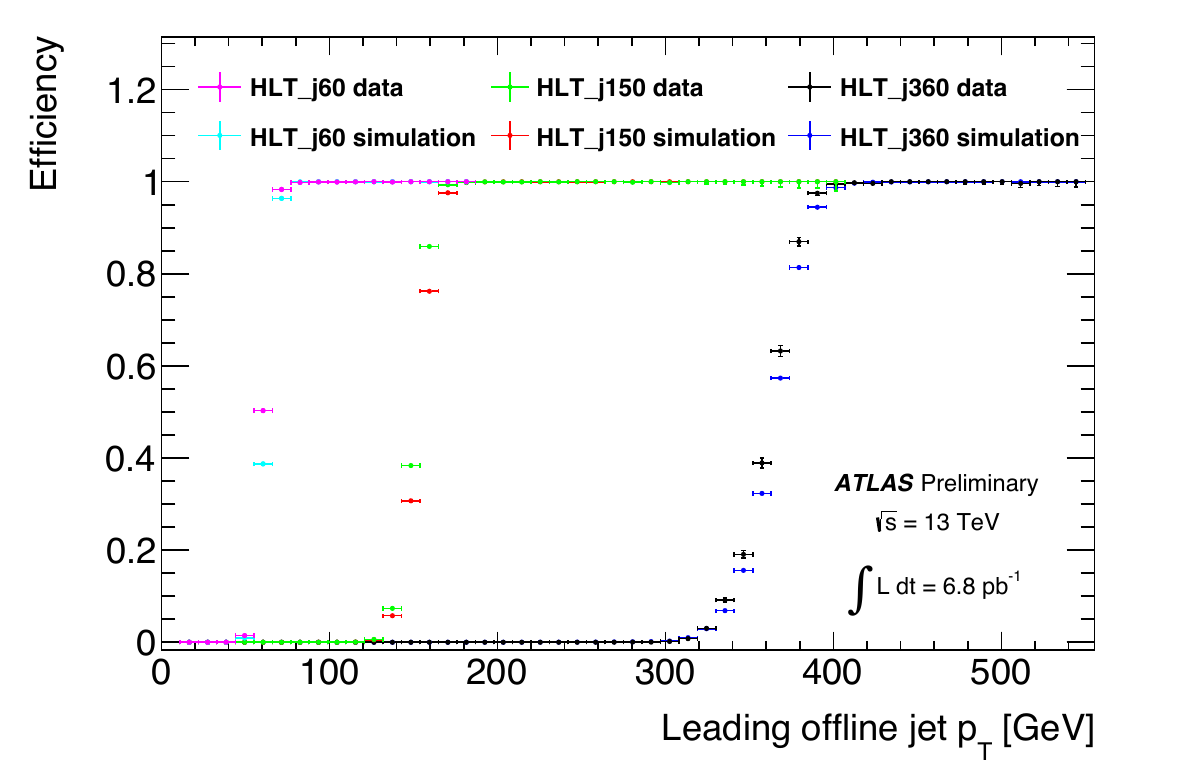}
\caption{Comparison of the HLT efficiency curves between the Run~II first week of data and the Monte Carlo simulation~\cite{JetTriggerPublic}.}
\label{fig:HLT}
\end{figure}

To assess the online behavior of the new software and the performance of the new jet energy corrections, the early data-taking was crucial. Shown in Fig.~\ref{fig:HLT} are the efficiency curves of some HLT jet triggers, for a wide range of threshold values, using the first week of stable beam data of Run~II~\cite{JetTriggerPublic}. The plateau regions for the data all reach 100\% efficiency and agree with those of the Monte Carlo simulations, showing that the new code kept a good performance in that respect. Also, an inspection of the figure reveals that the turn-on regions are nearly centered on their threshold value. A more quantitative description would require further work, but already this qualitative observation gives credence that the new jet energy calibrations are performing well.

\section{Conclusion}
The Run~II jet trigger code had to be completely rewritten to adapt to the new running conditions of the LHC. This new software relies on the improvement of the trigger, both in design and in hardware. To improve upon the Run~I performance, the new code can now apply some of the jet energy corrections as they are applied offline. All of these updates show great performance.

In the future, to further the improvements to the jet trigger, adding supplementary offline-inspired jet energy corrections is planned. These new corrections would mainly improve the resolution of the jet energy measured online.


\end{document}